\documentclass[usenatbib]{mn2e}
\usepackage{graphicx,graphics,amsmath}

\usepackage{float}
\usepackage[caption = false]{subfig}
\usepackage{mathptmx}
\usepackage{xcolor}
\usepackage[T1]{fontenc}
\DeclareRobustCommand{\VAN}[3]{#2}
\let\VANthebibliography\thebibliography
\def\thebibliography{\DeclareRobustCommand{\VAN}[3]{##3}\VANthebibliography}
\title[GR shock in neutron stars]{General relativistic shocks in connection with neutron stars}

\author[R. Mallick \& Anshuman]{Ritam Mallick$^{1}$\thanks{E-mail: mallick@iiserb.ac.in}
\& Anshuman$^{1}$\thanks{E-mail: anshuman18@iiserb.ac.in}
\\
$^{1}$Indian Institute of Science Education and Research, Bhopal, India}
\pubyear{2021}

\begin{document}
\label{firstpage}
\maketitle            
             
\begin{abstract} 
Astrophysical shocks are very common and are interesting as they are responsible for particle acceleration in supernova, blazers and in neutron stars. In this work, we study general relativistic shocks from the frame of the front. We derive the jump conditions and the Taub adiabat equation for both the space-like and time-like shocks. We solve these equations in a neutron star system where the shock is followed by a combustion front which is deconfining hadronic matter to quark matter. The maximum mass of the daughter quark star (generated from the combustion of the parent neutron star) is consistent with the maximum mass limit for the EoS sequence. We find that matter velocities for GR shocks under suitable condition can break the speed of light limit indicating very fast combustion process. Also, the matter velocities implies that for space-like shocks the combustion process is most probably a deflagration and for time-like shocks it is a detonation and can even proceed with velocities that are super-luminous. 
\end{abstract}

\maketitle

\section{Introduction}

Shock waves are generated by the discontinuous change in the thermodynamic variables of the system which travels faster than the local speed of sound. They are usually caused by the sudden compression or expansion of matter. Shocks are very common in astrophysics as they are seen in supernova, binary neutron star collision and are  associated with stellar winds, supernovae remnants, radio
jets, accretion on to compact objects, and phase transition (PT)
in neutron stars (NSs) \citep{amir,sironi,joel,lundman,amano}. When the shock is radiation dominated they are also believed to be the sources of non thermal photons, cosmic-rays and neutrinos \citep{zeldovich,ronaldo,tavani,jones}. In the last few decades renewed interest has been generated regarding shock wave not only in astrophysics and cosmology but also in high energy physics \citep{elth,carrus,smoller,kons,blandford,bland}.

Non-relativistic shock wave was known for a long time and is also very well studied \citep{landau,bethe,hirs,raizer,richard,shapiro}. The relativistic theory of shock wave started with Taub \citep{taub} and Landau \citep{landau}, where the rankine-hugoniot (RH) jump condition (basically energy-momentum and mass flux conservation equations) are solved to derive a single equation known as Taub adiabat (TA) equation connecting thermodynamical matter properties of the upstream and downstream matter.
Relativistic shocks were further analysed rigorously by Lichnerowicz \citep{lichnero,thorne,ran} assuming matter as an ideal fluid having an infinite conductivity. Since then a number of astrophysical problem was addressed in the literature \citep{colgate,anile,toshi,mangano,padua,eliz,gs,hoover}. In all the calculation the velocity of the shock is less than the speed of light (c). The shock surface or front forms a hyper-surface which is space-like (SL) which is to say that the shock hyper-surface has a SL normal vector. However, this is not only the case and it was pointed out by Csernai \citep{csernai} that that the normal to the hyper-surface can also be time-like (TL). He argued that a system undergoing rapid rarefaction, can generate bubbles at different spatial point not causally connected. The hyper-surface then becomes TL with bubble formation and growth.  
As an example they pointed out that as supercooled quark-gluon plasma fireball expands it cools and gets hadronized in heavy-ion collision. 
Various works then analyzed these results mainly in the context of heavy-ion collision \citep{gorenstein,rosenhauer,gyulassy}. 

In shock wave usually the matter properties behind and in front of the shocks are same (satisfying same equation of state (EoS)). There is only compression and rarefaction of matter due to shocks. However,
if the initial and final state of matter across the shock discontinuity does not belong to the same EoS, then we also have an combustion. The initial and the final state then does not lie in the same curve because of the difference of chemical energy. Thus the TA equation is called combustion or Chapman-jouget adiabat (CA). CA are particularly interesting when we study astrophysical PT \citep{ritam-irfan,ritam-shailendra-rana} in connection with NSs. After the proposition of the conjecture that strange quark matter (SQM) is the most stable state of matter at high density \citep{itoh,witten,bodmer}, there has been a renewed interest in the astrophysical community to test the theory. And the best laboratory to test this theory is the core of the NSs where central density can reach as high as $4-6$ times nuclear saturation density. Such densities are ideal for conversion from hadronic matter (HM) to quark matter (QM).  

One of the model for the PT process involves shock induced PT whose kinematics can be studied using RH conditions and TA equation \citep{bhattacharyya1,igor,schramm,ritam-amit,ritam-shailendra-rana}. They assume that a shocks wave initiates a combustion where the upstream matter is HM and the downstream matter is the QM. Then they solve the RH equations or the CA equation to obtain the downstream matter properties along with the matter velocities at either side of the front. Comparing the matter velocities with local speed of sound one can identify the combustion process. Most of the calculation done previously assume relativistic RH condition. However, if we are to extend them to analyze the problems in astrophysics we need to extend the relativistic results to general relativistic (GR) regime. That is the main aim of this work. In Section II we give the formalism of extending the calculation from relativistic to general relativistic regime. Section III discusses and compares the results of the relativistic and general relativistic case and also a simple application of the result for neutron star. We summarize and discuss our results in section IV.

\section{Formalism}

We begin with the assumption that the matter is flowing along the radial direction and the shock front is perpendicular to it. The width of the discontinuity is negligible in comparison to the actual system and is assumed to be a single discontinuity across which the flow variables are discontinuous. Denoting the two sides of the shock discontinuity as “a” and “b”, the difference of a thermodynamic quantity (Q) across the front is given by [Q] =$Q_{a}$-$Q_{b}$. The discontinuous surface is denoted by $\Sigma$ having a unit normal vector, $\wedge^{\mu}$, in space-time. The shock discontinuity can be SL or TL according to  the normalization condition given by
\begin{align}
    \wedge^{\mu}\wedge_{\mu} & = -1 && \text{(For TL Hyper-surface($\Sigma$))}& \nonumber \\
                             & = +1 && \text{(For SL
                             Hyper-surface($\Sigma$))}\nonumber
\end{align}

\begin{figure}
\hspace{0.5cm}
  \includegraphics[scale=0.8]{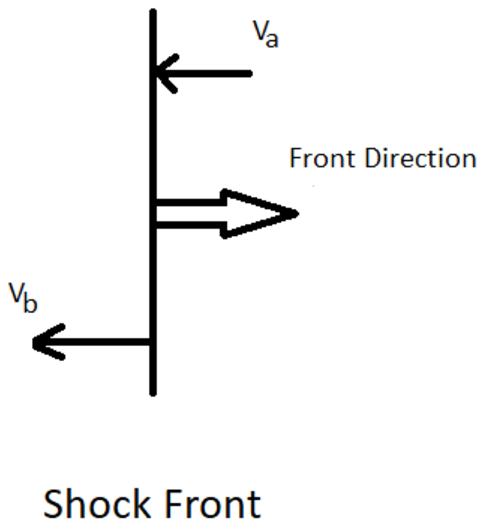}
  \caption{Schematic diagram of a moving shock discontinuity. The shock front is moving from left to right. $v_a$ and $v_b$ are the matter velocities on either side of the front.}
\end{figure}

\subsection{Relativistic Shock Waves}
The relativistic fluid dynamic equations is constructed from the energy-momentum tensor ($T^{\mu\nu}$) in Minkowski ST, where $T^{00}$ is the energy density, $T^{0\alpha}$ is the energy flux  and $T^{\alpha\beta}$ the momentum flux. 
 

The line element in flat ST is given by

\begin{equation}
    ds^2= \eta_{\mu\nu}dx^{\mu}dx^{\nu}=-dt^2 + dr^2 + r^{2}d\Omega^{2},
\end{equation}
where $\eta_{\mu\nu} $ is flat space-time metric which in spherical polar coordinates is given by 

\begin{equation*}
\eta_{\mu\nu} =
\begin{pmatrix}
- 1 & 0 & 0 & 0\\
 0  &+1  & 0 & 0\\
 0   & 0 & r^{2} & 0\\
 0   & 0 & 0 & r^{2}\sin ^2\theta 
\end{pmatrix}
\end{equation*}

Considering a perfect fluid (i.e, fluid with no viscosity or heat conductivity), the stress-energy tensor is given by
\begin{equation}
    T^{\mu\nu} = wu^{\mu}u^{\nu} + pg^{\mu\nu}
\end{equation}
where, $w$ (the enthalpy) is equal to the $e$ (energy density) + $p$ (pressure), $u^{\mu}=(\gamma,\gamma v,0,0)$ is the four-velocity vector with $\gamma = \frac{1}{\sqrt{1 - v^{2}}}$ and its norm is given by $g_{\mu\nu}u^{\mu}u^{\nu} = -1$.

The conservation of energy-momentum tensor ($\partial_{\mu}T^{\mu\nu}=0$) and mass-flux ($\partial_{\mu}J^{\mu}=0$) in the rest frame of the front becomes the RH jump condition across the discontinuity
\begin{align}
    T_{a}^{\mu\nu}\wedge_{\nu} & = T_{b}^{\mu\nu}\wedge_{\nu} \\
    n_{a}u_{a}^{\mu}\wedge_{\mu} &= n_{b}u_{b}^{\mu}\wedge_{\mu}.
\end{align}

\begin{itemize}
    \item For SL discontinuity, the normal vector (NV) is given by 
    \begin{equation}
        \Lambda^{\mu} = (0,1,0,0) \nonumber
    \end{equation}
    Therefore, the RH jump conditions becomes 
    \begin{align}
      T_{a}^{01} & =  T_{b}^{01} \nonumber\\
      \Rightarrow w_{a}\gamma_{a}^{2}v_{a} & = w_{b}\gamma_{b}^{2}v_{b} \\
      T_{a}^{11} & =  T_{b}^{11} \nonumber \\
      \Rightarrow w_{a}\gamma_{a}^{2}v_{a}^{2} + P_{a} & = w_{b}\gamma_{b}^{2}v_{b}^{2} + P_{b} \\
      n_{a}u_{a}^{1} & = n_{b}u_{b}^{1} \nonumber \\
      \Rightarrow n_{a}v_{a}\gamma_{a} & = n_{b}v_{b}\gamma_{b}
    \end{align}
    
    \item For TL discontinuity, the NV is given by
    \begin{equation}
        \Lambda^{\mu} = (1,0,0,0). \nonumber
    \end{equation}
    and the RH jump conditions are given as 
        \begin{align}
      T_{a}^{10} & =  T_{b}^{10} \nonumber\\
      \Rightarrow w_{a}\gamma_{a}^{2}v_{a} & = w_{b}\gamma_{b}^{2}v_{b} \\
      T_{a}^{00} & =  T_{b}^{00} \nonumber \\
      \Rightarrow w_{a}\gamma_{a}^{2} - P_{a} & = w_{b}\gamma_{b}^{2} - P_{b}\\ 
      n_{a}u_{a}^{0} & = n_{b}u_{b}^{0} \nonumber \\
      \Rightarrow n_{a}\gamma_{a} & = n_{b}\gamma_{b}
    \end{align}
\end{itemize}

where $\gamma_i = \frac{1}{(1-v_i^2/c^2)^{1/2}} = \frac{1}{(1-v_i^2)^{1/2}}$, where $i=a, b$. The above equations can be further solved to derive the TA/CA equation both for the SL and TL case and is given by
\begin{equation}
   \Bigg(\frac{w_{a}^2}{n_{a}^{2}} - \frac{w_{b}^2}{n_{b}^{2}}\Bigg) + (p_{b} - p_{a})\Bigg(\frac{w_{a}}{n_{a}^{2}} + \frac{w_{b}}{n_{b}^{2}}\Bigg) = 0. \\
\end{equation}
Interestingly, for the relativistic case the TA/CA equation for both SL and TL comes out to be the same. 

The TA equation does not have the velocity terms, however knowing the matter properties on both sides of the front, the velocity of the matter of both phases can be calculated. They are given as

\begin{itemize}
 \item SL velocities
 \begin{align}
    v_a = \sqrt{\frac{(p_b - p_a)(e_b + p_a)}{(e_b - e_a)(e_a + p_b)}} \nonumber \\
    v_b = \sqrt{\frac{(p_b - p_a)(e_a + p_b)}{(e_b - e_a)(e_b + p_a)}} \nonumber 
\end{align}

\item TL velocities
\begin{align}
    v_a = \sqrt{\frac{(e_b - e_a)(e_a + p_b)}{(p_b - p_a)(e_b + p_a)}} \nonumber\\
    v_b = \sqrt{\frac{(e_b - e_a)(e_b + p_a)}{(p_b - p_a)(e_a + p_b)}} \nonumber 
\end{align}
\end{itemize}

\subsection{General Relativistic Shock Waves}
For general relativistic shock the line element is given by (for a spherically symmetric space-time (ST))
\begin{equation}
    ds^2= g_{\mu\nu}dx^{\mu}dx^{\nu}=- e^{2\phi(r)}dt^2 + e^{2\Lambda(r)}dr^2 + r^{2}d\Omega^{2}
\end{equation}
where $g_{\mu\nu}$ is general spherically symmetric metric given by
\begin{equation*}
g_{\mu\nu} =
\begin{pmatrix}
- e^{2\phi(r)} & 0 & 0 & 0\\
 0   & e^{2\Lambda(r)}  & 0 & 0\\
 0   & 0 & r^{2} & 0\\
 0   & 0 & 0 & r^{2}\sin ^2(\theta )
\end{pmatrix}
\end{equation*}

Although the form of the energy-momentum tensor remains the same, the four velocity in curved ST is given by
\begin{equation}
    u^{\mu} = \dfrac{dx^{\mu}}{ d\tau}  = \dfrac{dx^{\mu}dt}{dt d\tau}\nonumber 
    \end{equation}    
where, the four position vector in spherical polar coordinate is $x^{\mu}=(t,r,\theta,\Phi)$ and the norm of $u^\mu$ is given by $g_{\mu\nu}u^\mu u^\nu = -1 $.

\par Considering the front to be moving along the radial direction the four-velocity of fluid particles is given by
\begin{equation}
    u^{\mu} = \gamma_g(1,v_r,0,0)
\end{equation}
where, $\theta $  and $\Phi =$ constant, $\gamma_g=\frac{1}{[e^{2\phi}-e^{2\Lambda}v_r^2]^{1/2}}$ and $v_r=\dfrac{dr}{dt}$ is radial velocity.


Following the same normalization conditions (as done for relativistic treatment), the NV for SL shock is
    \begin{equation}
        \Lambda^{\mu} = (0,\frac{1}{\sqrt{g_{11}}},0,0)
    \end{equation}
and the NV for TL discontinuity is
    \begin{equation}
        \Lambda^{\mu} = (\frac{1}{\sqrt{-g_{00}}},0,0,0) 
    \end{equation}
where, $g_{00} = - e^{2\phi(r)}$ and $g_{11} = e^{2\Lambda(r)} $ are the metric elements.

Having defined our ST metric and the shock normal, the jump condition for GR shocks is given by 

\begin{itemize}
    \item For SL shock
    \par the energy-flux conservation jump condition is 
    \begin{align}
      T_{a}^{01} & =  T_{b}^{01} \nonumber\\
      \Rightarrow w_{a}\gamma_{ga}^2 v_{ra} & = w_{b}\gamma_{gb}^2 v_{rb} 
      \end{align}
      \par the momentum-flux conservation jump condition becomes
      \begin{align}
      T_{a}^{11} & =  T_{b}^{11} \nonumber \\
      \Rightarrow w_{a}\gamma_{ga}^2 v_{ra}^2 + \frac{p_a}{e^{2\Lambda_a}} & = w_{b}\gamma_{gb}^2 v_{rb}^2 + \frac{p_b}{e^{2\Lambda_b}}
      \end{align}
      \par and the particle-flux conservation is given by \\
      \begin{equation}
          n_{a}u_{a}^{1} = n_{b}u_{b}^{1} \nonumber \\     
      \end{equation}
      \begin{align}
      \Rightarrow n_a\gamma_{ga}v_{ra} = n_b\gamma_{gb}v_{rb} =j      
    \end{align}
    \par Using eqn 17 and 18 the particle-current density (j) is given by
    \begin{align}
        j^2 & = (\frac{p_b}{e^{2\Lambda_b}}-\frac{p_a}{e^{2\Lambda_a}})\frac{1}{(w_aV_a^{2}-w_bV_b^{2})}
    \end{align}
    \par where, $V_a=\frac{1}{n_a}$ and $V_b=\frac{1}{n_b}$.
    \item For TL shocks
    \par \par the energy-flux conservation
    \begin{align}
      T_{a}^{10} & =  T_{b}^{10} \nonumber\\
      \Rightarrow w_a\gamma_{ga}^2 v_{ra} & = w_b\gamma_{gb}^2 v_{rb} 
      \end{align}
      \par the momentum-flux conservation
      \begin{align}
      T_{a}^{00} & =  T_{b}^{00} \nonumber\\
       \Rightarrow w_a\gamma_{ga}^2 - \frac{p_a}{e^{2\phi_a}} & = w_b\gamma_{gb}^2 - \frac{p_b}{e^{2\phi_b}}
      \end{align}
      \par particle-flux conservation \\
      \begin{equation}
          n_{a}u_{a}^{0} = n_{b}u_{b}^{0} \nonumber \\
      \end{equation}
     \begin{align}
      \Rightarrow n_a\gamma_{ga} =  n_b\gamma_{gb} = j
    \end{align}
    \par where j$^2$ is given by
    \begin{align}
         j^2 & = (\frac{p_a}{e^{2\phi_a}}-\frac{p_b}{e^{2\phi_b}}) \frac{1}{(w_aV_a^{2}-w_bV_b^{2})}
    \end{align}
\end{itemize}

Once the Jump condition are derived, we then proceed to derive the corresponding TA/CA for the SL and TL shocks.

\begin{itemize}
    \item For SL Shocks, using eqn 18, $v_{ra}$ and $v_{rb}$ becomes \\
\end{itemize}
\begin{align}
    v_{ra}^2 = \frac{e^{2\phi_a}}{\frac{1}{j^2V_a^2}+e^{2\Lambda_a}} \\
    v_{rb}^2 = \frac{e^{2\phi_b}}{\frac{1}{j^2V_b^2}+e^{2\Lambda_b}}
\end{align}
Squaring eqn 16 and using above equations (i.e, the value of $v_{ra}^{2}$ and $v_{rb}^{2}$ with $j^2$), the TA/CA equation for SL shocks becomes
\begin{equation}
    (w_a^{2}\gamma_{ga}^4 v_{ra}^{2} - w_b^{2}\gamma_{gb}^4 v_{rb}^{2} ) = 0 \nonumber 
\end{equation}
\begin{align}
    \Rightarrow &\left[\frac{p_b}{e^{2\Lambda_b}} - \frac{p_a}{e^{2\Lambda_a}}\right]\left[\frac{w_a^2V_a^4e^{2\Lambda_a}}{e^{2\phi_a}} - \frac{w_b^2V_b^4e^{2\Lambda_b}}{e^{2\phi_b}}\right]\nonumber\\
    &+ (w_aV_a^2-w_bV_b^2)\left[ \frac{w_a^2V_a^2}{e^{2\phi_a}} - \frac{w_b^2V_b^2}{e^{2\phi_b}}\right] = 0.
\end{align}
     
\begin{itemize}
    \item Similarly, we can derive the TA/CA for TL shock and is given by
\end{itemize}
\begin{equation}
    (w_a^{2}\gamma_{ga}^4 v_{ra}^{2} - w_b^{2}\gamma_{gb}^4 v_{rb}^{2} ) = 0 \nonumber 
\end{equation}
\begin{align}
    \Rightarrow &\left[\frac{p_b}{e^{2\phi_b}} - \frac{p_a}{e^{2\phi_a}} \right]\left[\frac{w_a^2V_a^4e^{2\phi_a}}{e^{2\Lambda_a}} - \frac{w_b^2V_b^4e^{2\phi_b}}{e^{2\Lambda_b}}\right] \nonumber \\
   & + (w_aV_a^2-w_bV_b^2)\left[  \frac{w_a^2V_a^2}{e^{2\Lambda_a}}-\frac{w_b^2V_b^2}{e^{2\Lambda_b}}\right] = 0
\end{align}

It is interesting to note that for the GR shocks the TA/CA equation are different for the SL and TL shocks unlike the relativistic case where they were same. The matter velocities of the two phases in terms of the thermodynamic variables are given by \\

SL shocks \\
\begin{align}
    {v_a} &= \sqrt{\frac{A_2B_1B_2w_a^{2} - a_{11}+b_{11}}{2c_{11}}} \\
    {v_b} &= \frac{{v_a}B_1(A_2 w_a + A_1 w_b  - \frac{b_{11}}{w_a B_1B_2} )}{2A_1[B_2p_a + B_1\epsilon_b]}
\end{align}
where, we have defined 
 \begin{align*}
     A_1 & = e^{2\phi_a} \text{,}
     A_2  = e^{2\phi_b} \text{,}
     B_1  = e^{2\Lambda_a} \text{,}
     B_2  = e^{2\Lambda_b}\\
     w_a & = (p_a + \epsilon_a) \text{,}
      w_b  = (p_b + \epsilon_b)\\
     a_{11} &= A_1\big[B_1B_2(\epsilon_a - p_a)(\epsilon_b - p_b ) + 2(B_2^{2}p_a\epsilon_a +B_1^{2}p_b\epsilon_b)]\\
     b_{11} &= w_a\sqrt{B_1B_2}\sqrt{B_1B_2(A_1^{2}w_b^{2}+A_2^{2}w_a^{2}) - 2A_2a_{11}}\\
     c_{11} &= B_1(B_1p_b + B_2\epsilon_a)(B_2\epsilon_a - B_1\epsilon_b) \\
 \end{align*}
 
TL shocks\\

\begin{align}
    {v_a} &= \sqrt{\frac{A_1(a_{21} - b_{21})}{2c_{21}}} \\
    {v_b} &= \frac{{v_a}A_2(B_2 w_a + B_1 w_b  + \frac{b_{21}}{w_a A_1A_2} )}{2B_2[A_1p_b + A_2\epsilon_a]}
\end{align}
 where, 
 \begin{align*}
     A_1 & = e^{2\phi_a} \text{,}
     A_2  = e^{2\phi_b} \text{,}
     B_1  = e^{2\Lambda_a} \text{,}
     B_2  = e^{2\Lambda_b}\\
     w_a & = (p_a + \epsilon_a) \text{,}
      w_b  = (p_b + \epsilon_b)\\
     a_{21} &= A_1A_2\big[B_2 w_a^{2} - B_1( \epsilon_b - p_b)( \epsilon_a - p_a)] - 2B_1(A_2^{2}p_a\epsilon_a +A_1^{2}p_b\epsilon_b)\\
     b_{21} &= w_a\sqrt{A_1A_2}\sqrt{A_1A_2(B_1^{2}w_b^{2}-B_2^{2}w_a^{2}) + 2B_2a_{21}}\\
     c_{21} &= B_1^{2}(A_2p_a + A_1\epsilon_b)(A_2p_a - A_1p_b) \\
 \end{align*}
The jump conditions, TA/CA equations and the matter velocities of the GR shocks reduces to the corresponding relativistic shocks (both for SL and TL) if either the metric potentials becomes zero
\begin{equation*}
     \Rightarrow \Lambda_a = \phi_a = \Lambda_b = \phi_b = 0 \\
     \end{equation*}
or if the metric potentials on either side of the front becomes equal, i.e. \\
 \begin{equation*}
     \Lambda_a = \Lambda_b = \Lambda; ~ \phi_a = \phi_b = \phi. \\
 \end{equation*}

\section{Results}

We assume that the shock front is accompanied with a combustion front which is altering the matter properties on the other side of the front. Therefore, we need two equation of state to describe matter properties on either side of the front. The initial matter EoS is most probably a zero temperature hadronic EoS as we are considering cold matter. 
We also assume that the final burnt matter EoS is of quark matter as the model system that we will study in this work is a neutron star (NS) where hadronic matter is undergoing a deconfinment to quark matter due to the shock induced burning.
Although, the final burnt QM can have finite temperature the temperature is less than $10$ MeV \citep{ritam-slow-burn}. Such low temperature has negligible effect of the EoS of quark matter \citep{igor,ritam-amit}.
For the hadronic phase, we adopt a relativistic mean-field EoS with PLZ parameter setting \citep{serot,glen,reinhard}.
The EoS of the quark matter is modelled after the MIT
bag model \citep{chodos} having u, d, and s quarks with mass $5$, $10$ and $80$ MeV. The bag
pressures is assumed to be $B^{1/4} = 160$ MeV and the quark interaction term is taken to be $0.6$ \citep{alford}.

It is possible to deduce the combustion process associated with a shock. They are categorized comparing the matter velocities and speed of sound on either side of the front like fast combustion or slow combustion. However, for relativistic shocks, the matter velocities are usually comparable to speed of light or even can be super-luminous. There can be even situation where the matter velocities becomes imaginary and non-physical. Starting with a given a initial state and solving for the final state one can then categorize different physical and non-physical region in an $\epsilon-p$ diagram, depending on the matter velocities. The two important velocity limits are \\
1. Matter velocities reaches the velocity of light ($v=c=1)$\\
2. Matter velocities becomes imaginary.
The condition for which the matter velocities of relativistic SL shocks reaches the speed of light is given by
\begin{align*}
    v_a^{2} &= 1 = v_b^{2}  \\
    \Rightarrow p_b &= \epsilon_b + p_a - \epsilon_a  
\end{align*}
 And the condition for which they become imaginary are 
\begin{align*}
    v_a^{2} &= -ve = v_b^{2} \\
    \Rightarrow p_b &<  p_a  \textrm{  \&  } \epsilon_b > \epsilon_a \\ \quad \textbf{    $   or   $   } \quad \\
      \Rightarrow p_b &>  p_a  \text{ \&} \epsilon_b < \epsilon_a
\end{align*}
It is interesting to note that for relativistic shocks the above two condition for TL shocks are same because $v_a^{sl} = \frac{1}{v_a^{tl}}$ and $v_b^{sl} = \frac{1}{v_b^{tl}}$. \\

However, for GR shocks the conditions for SL and TL shocks are different. For the GR SL shocks the condition for luminous matter velocities are \\
a) condition for $v_a \approx 1$ is
\begin{align*}
    p_b &=\frac{{B2} \left[p_a h_1 +{A_1} {B_1} h_2 + {A_2}{B_1}^2 h_3 + {B_1}^2 {\epsilon_a} ({B_2} {\epsilon _a} - {B_1} {\epsilon_b})\right]}{{B_1} \left(h_1 + {A_1} {B_1} h_4 + {B_1}^2 ({B_1}{\epsilon_b} - {B_2}{\epsilon_a})\right)}.
\end{align*}
b) and the condition for $v_b \approx 1$ is
 \begin{align*}
p_b &= \frac{{B_2} (- h {A_1} {B_1}{\epsilon_b}- {A_1} {B_2}{p_a}+{A_2}{B_1}{p_a}+ h {A_2}{B_1} {\epsilon_a})}{{B_1} ({A_2}{B_1}-{A_1}{B_2})}.
\end{align*} 
where,
\begin{align*}
    h_1 & = {A_1}^2 ({B_1} {\epsilon_b}+{B_2}{p_a})\\
    h_2 & = - {B_1} {p_a} {\epsilon_b}+ {B_1} {\epsilon_a} {\epsilon _b} + 2 {B_2} {p_a} {\epsilon_a} \\
    h_3 & = - {p_a}^2 - 2 {p_a} {\epsilon_a} - {\epsilon_a}^2 \\
    h_4 & = - 2 {B_1} {\epsilon_b} - {B_2} {p_a} + {B_2} {\epsilon_a}
\end{align*}
The luminous matter velocity conditions for GR TL Shock are \\
a) the condition for $v_a \approx 1$ becomes
    \begin{align*}
 p_b &= \frac{{A_2}({A_2}G1+{A_1}G3)}{{A_1} \left( {A_1}^3 {\epsilon_b}+{A_1}^2 G4 + {A_1}{B_1} G5 +{A_2}{B_1}^2 {p_a}\right)}
\end{align*}
b) and the condition for $v_b \approx 1$ becomes
\begin{align*}
p_b &= \frac{h {A_2} [{A_1} ({B_1}{\epsilon_b}+{B_2} {p_a}-{B_2}{\epsilon_a})-{A_2}{B_1}{p_a}]}{{A_1} ({A_1}{B_2}-{A_2}{B_1})} 
\end{align*}
where,
    \begin{align*}
    G1 &= ({A_1}^2{\epsilon_a}^2+2{A_1}{B_1}{p_a}{\epsilon_a}+{B_1}^2 {p_a}^2) \\
    G2 &=  (-{B_1}{p_a} {\epsilon_b}+{B_1} {\epsilon_a} {\epsilon_b}-{B_2}{p_a}^2-2 {B_2} {p_a}{\epsilon_a} -{B_2}{\epsilon_a}^2)\\
    G3 &= -{A_1}^2 {\epsilon_a} {\epsilon_b}+{A_1}G2 + {B_1}^2{p_a}{\epsilon_b} \\
    G4 &= -({A_2}{\epsilon_a}-2 {B_1} {\epsilon_b})\\
    G5 &=  (-{A_2}{p_a}+{A_2}{\epsilon_a}+{B_1}{\epsilon_b})\\
    h & = Constant.\\ 
    \end{align*}
Similarly, the condition for which the matter velocities becomes imaginary  for GR SL shocks are (same for both $v_a$ and $v_b$) 

\begin{align*}
B_2\epsilon_a &< B_1\epsilon_b  \textrm{ \&  } (A_2B_1 B_2w_a^{2} + b_{11}) > a_{11} \\
or \\
B_2 \epsilon_a &> B_1 \epsilon_b  \textrm{ \&  } (A_2B_1 B_2w_a^{2} + b_{11}) < a_{11} \\
\end{align*}
 
For TL shocks, the condition comes out to be



\begin{align*}
A_2 p_a &< A_1 p_b  \textrm{  \&  } a_{21} > b_{21} \\
or \\
A_2 p_a &> A_1 p_b  \textrm{   \&  } a_{21} < b_{21}  
\end{align*}

\begin{figure*}
\centering
\includegraphics[width = 7.0in,height=2.70in]{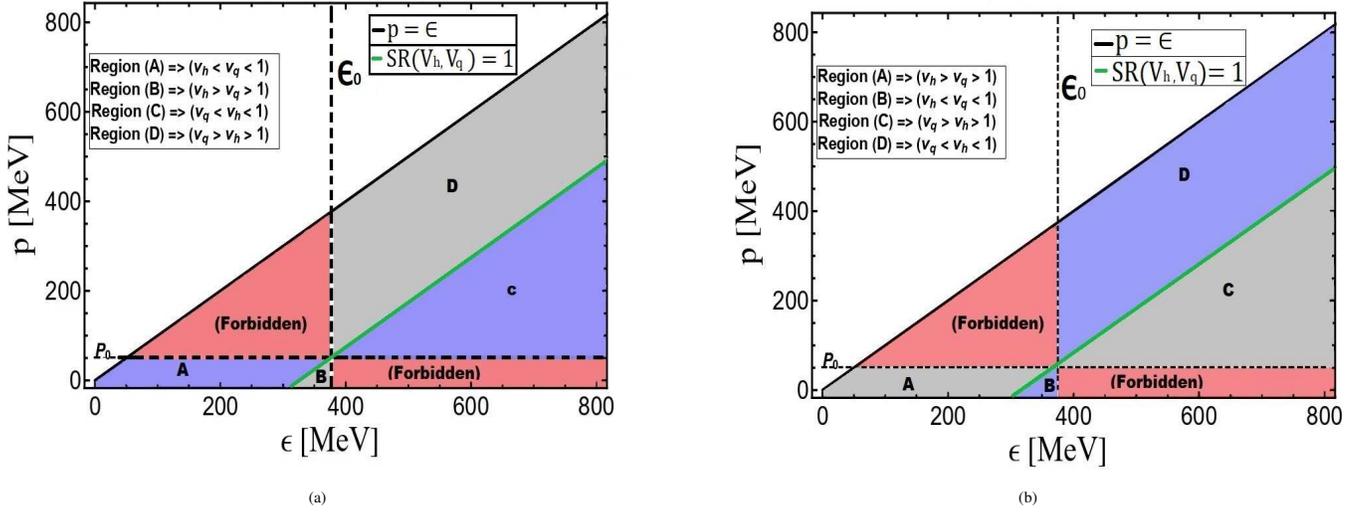}
\hspace{3.0cm} \scriptsize{(a)} \hspace{9.10cm} \scriptsize{(b)} 
\caption{Final states in the pressure (p) versus energy-density ($\epsilon$) curves are shown for shock induced transition from initial state ($\epsilon_0$= $\epsilon_a$,$p_0$ = $p_a$) for (a) SL shock and (b) TL shock in relativistic case. For SL shock, A and C regions are sub-luminal (blue), while B and D regions are super-luminal (gray). However, for TL Shock, B and D regions are sub-luminal (blue), but A and C regions are super-luminal (gray). Other regions (red) for both SL and TL shocks are unphysical (where velocities are imaginary).}

\end{figure*}

Starting with a given initial condition ($p_0,\epsilon_0$) we can find the region corresponding to different velocity conditions in the $\epsilon-p$ diagram. We start with the relativistic SL shocks whose different regions are shown in fig.(2(a)). The black solid line corresponds to the $p=\epsilon$ line which is the causality limit. The region above this line is the global forbidden region. We will characterize the region below this line according to the luminous and imaginary velocity conditions.
Region A and C are the regions where the matter velocities are less than the speed of light. However, there are also regions below the solid black line for which the matter velocities becomes super-luminous (region B and D). With the given initial condition if we are able to reach any point in the region as the final point then the matter velocities will break the speed of light limit.  There are also forbidden regions (marked in red) where the matter velocities with the given initial state becomes imaginary.

Similar region-plot is shown in fig(2(b)) for relativistic TL shocks. For this case the forbidden region remains the same however, the region of sub and super luminous matter velocities gets interchanged as the matter velocities for SL and TL shocks are inverse of each other. Previously, in the literature these regions have been mentioned earlier \citep{csernai,gorenstein} and they argue that spontaneous hadronization seen in heavy ion process can be explained by such super-luminous velocities \citep{gorenstein,rosenhauer}.

However, the uniqueness of our present work is in the fact that we have extrapolated this idea to GR shocks. The problem with relativistic shock is simple in comparison to GR shocks. To study relativistic shocks we do not need any particular system as the metric is always the same. However, this is not the case for GR treatment and we always need a particular system to get the metric potentials which vary from system to system. In the present work our model system is the interior of a neutron star (NS). We solve the TOV equation (eqn \citep{tov}) for a certain initial density and obtain the pressure and density profile as a function of radius. Knowing the pressure and density profile we also find the metric potential as a function of pressure (and thereby of radius). We follow the procedure for both the hadronic and quark EoS.

The region - plot for GR SL shocks is shown in fig(3(a)). The causality line remains the same. The initial conditions are given by $p_0,\epsilon_0$ and the different regions are shown with different colours. Analytical results are not possible for the GR shocks and we have solved our equations numerically to obtain the different regions. The corresponding $v_h=1$ and $v_q=1$ lines are shown by green and red line respectively. The forbidden region in the red (a small region) for Sl shocks. Region A and B are the two allowed region where matter velocities are either smaller the velocity of light or are very close to it. Region C and D are regions where matter velocities super-luminous. The region - plots for GR shock is different from SR case in the sense that the relation between pressure and energies are no longer linear. The regions are effected by the gravitational potentials i.e by the mass of the star. 

The corresponding TL GR shock regions are shown in fig(3(b)). However, unlike the SR case the matter velocities of the SL and TL GR shocks are not inverse to each other and therefore the TL GR regions are not exactly opposite of SL GR shock region. For GR TL shocks a huge region is forbidden where the matter velocities becomes imaginary. Region A and B are regions where matter velocities becomes super-luminous. Region C and D are regions where the matter velocities are less than speed of light. Also, there is a small window between the forbidden region where the matter velocities are finite. 
It should be clear that the region - plots differ with different initial conditions, especially for GR shocks.

\begin{figure*}
\centering
\includegraphics[width = 7.0in,height=2.70in]{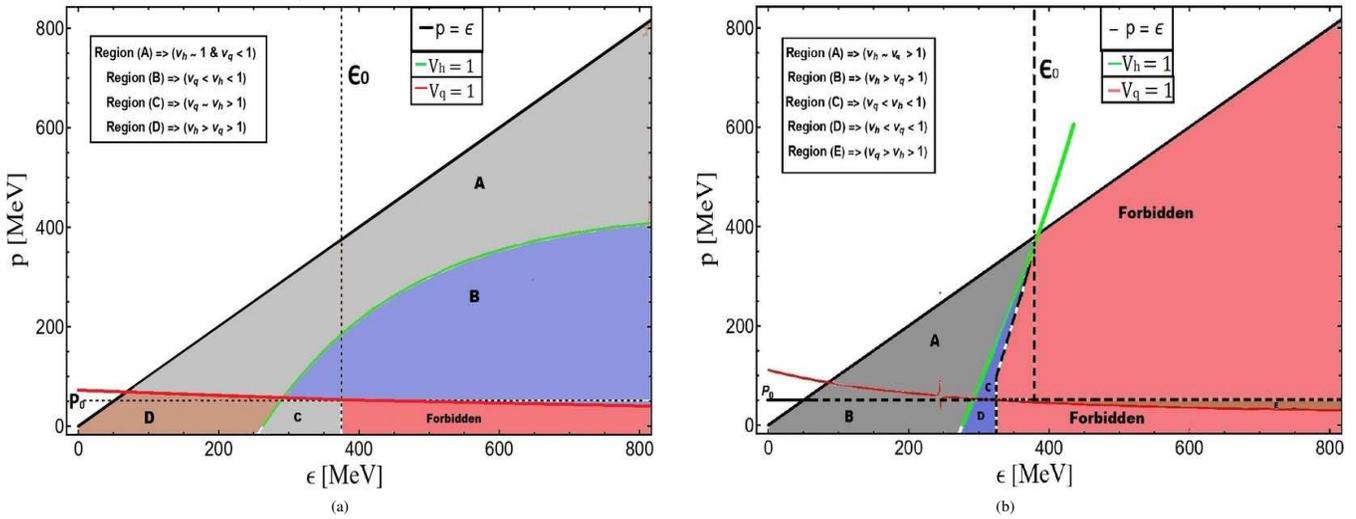}
\hspace{3.60cm} \scriptsize{(a)} \hspace{8.50cm} \scriptsize{(b)} 
\caption{Final states in the pressure (p) versus energy-density ($\epsilon$) curves are shown for shock induced transition from initial state ($\epsilon_0$= $\epsilon_a$,$p_0$ = $p_a$) for (a) SL shocks and (b) TL shocks in general relativistic case. For SL shock, B region is sub-luminal (blue) while in region A velocities are less than or equals to luminal velocity but C and D regions are super-luminal (gray). However, for TL Shock, C and D regions are sub-luminal (blue), but A,B and E regions are super-luminal. Other regions(red) for both SL and TL shocks are unphysical (where velocities become imaginary).}

\end{figure*}

\subsection{Relativistic and General Relativistic Combustion Adiabat}
The region - plots are important and exciting from the physics point of view in the sense that it give rise to super-luminous velocities which can have important physical applications in connections to astrophysical shocks. However, in the present problem we will focus on GR CA and study the combustion of NS to QS whose relativistic study has been done earlier \citep{ritam-irfan,ritam-shailendra-rana}. The relativistic CA was used to predict and constraint the maximum mass of the daughter QS which results from the combustion of parent NS. In the present work we extend the study and include GR corrections.

\begin{figure*}
	\centering
		\includegraphics[width = 3.45in,height=2.5in]{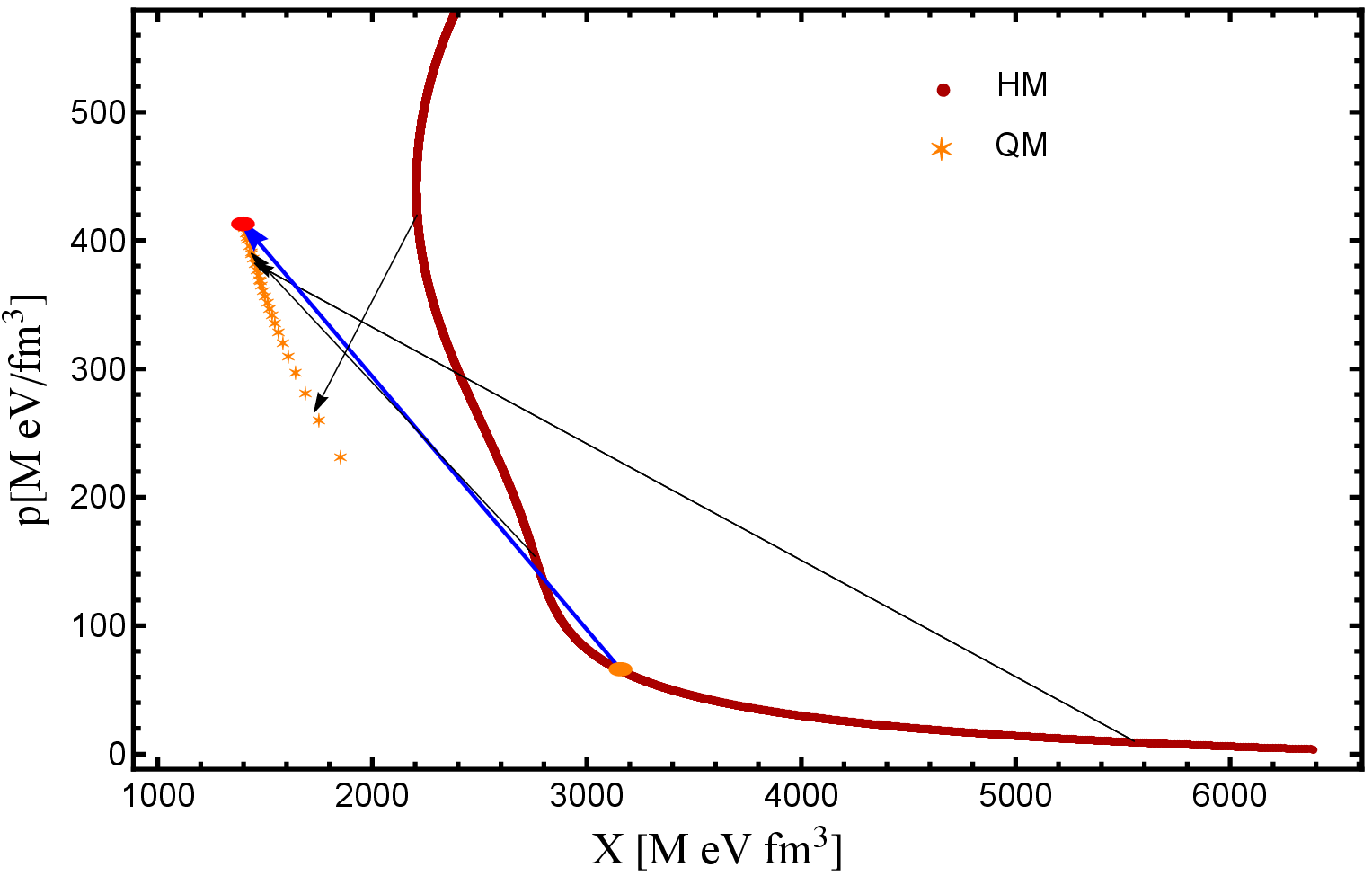}
		\includegraphics[width = 3.45in,height=2.5in]{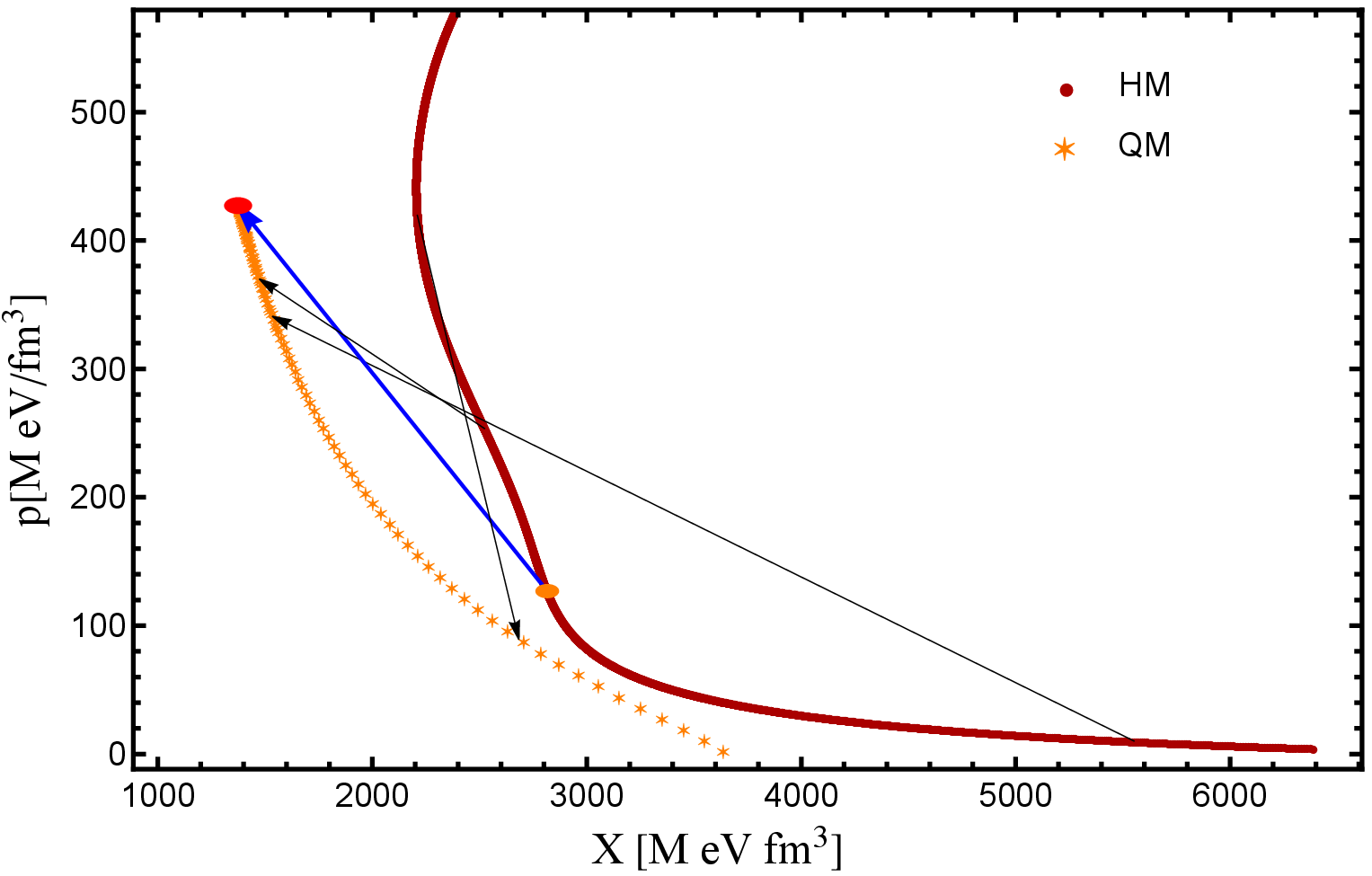}
			\hspace{0.5cm} \scriptsize{(a)} \hspace{9.50 cm} \scriptsize{(b)}
	\hspace{3.0cm}	\includegraphics[width = 3.45in,height=2.5in]{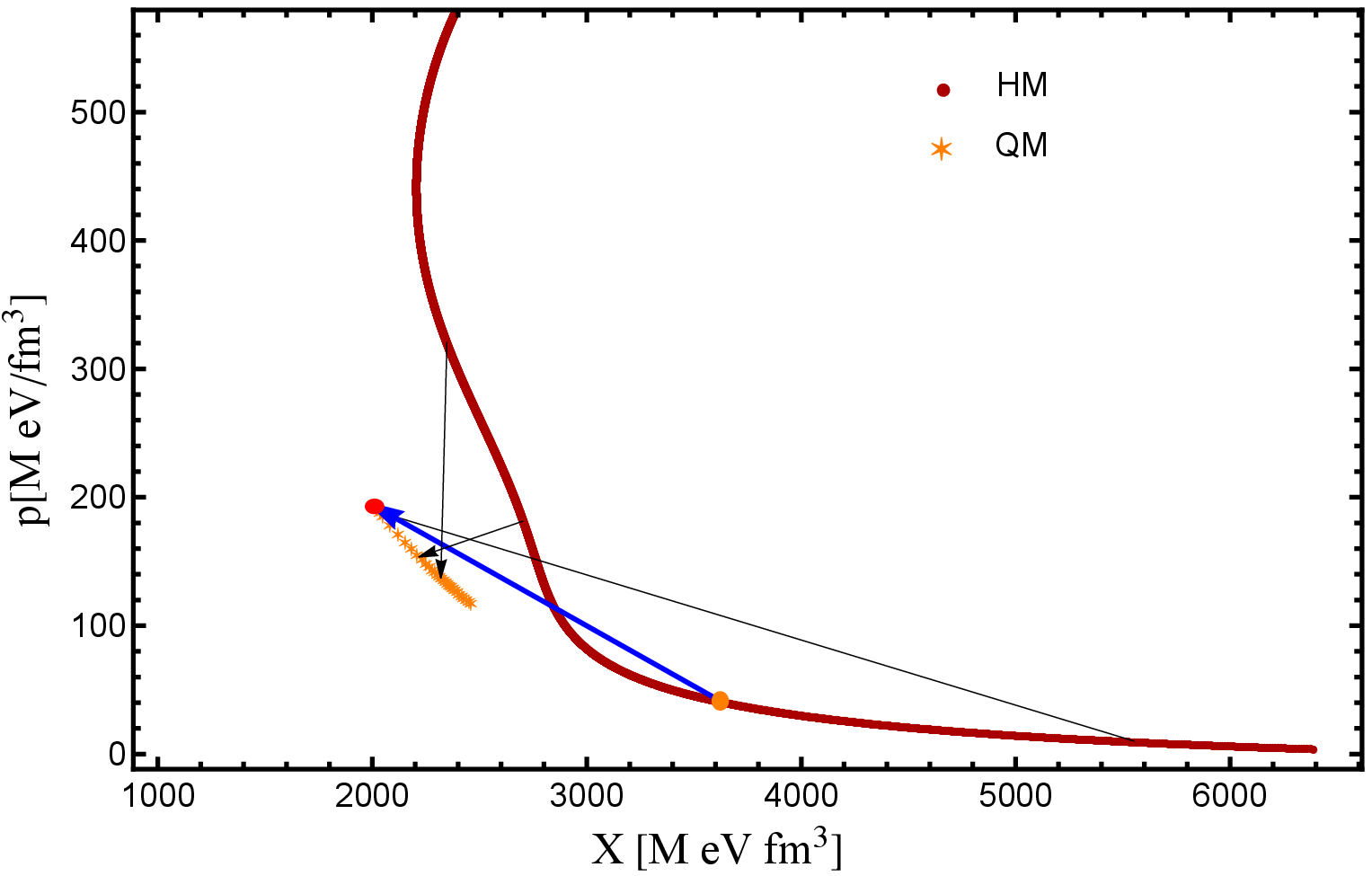}
		\hspace{.6cm} \\
		\scriptsize{(c)}
\caption{(a) General Relativistic CA (p versus x) curve (SL) (b) General Relativistic CA (p versus x) curve (TL) (c) Relativistic CA (p versus x) curve (SL/TL), shows HM EoS (red-solid line with circle) with their corresponding burnt state QM (orange-star). The red-dot on QM EoS indicate the maximum on the burnt QM EoS corresponding to orange-dot on HM EoS. In the low pressure arrows indicate the regular jump from HM EoS to its burnt state, but as pressure become higher, arrow indicate the retracing nature of QM curves.}

\end{figure*}

The initial state or the upstream quantities are the input (here the HM EoS). Also, the final EoS of the burnt or downstream state is also known (the QM EoS). The CA are used to calculate the the corresponding state in the downstream matter for a given initial state. As the EoS are different the upstream and downstream point lie on different curve. We can plot the curves for the HM and its corresponding QM in the $X-p$ plane. The initial point lie in the red HM curve and for a given initial state, solving the CA equation we obtain a point lying in the burnt QM (orange curve). The lines connecting the initial and the final point is known as the "Rayleigh line" whose slope is proportional to $(\gamma_n v_n)^2$. The CA curve for SL and TL relativistic shock is shown in fig(4(c)). Initially, there is a compression due to the shock which means a initial point with smaller pressure corresponds to a final point with larger pressure. As we go up the HM curve the slope of RL decreases however, both the upstream and downstream curve rises. But, there is a maximum point on the burnt trajectory (or downstream curve) beyond which if we go along the HM curve, the downstream curve comes down and retrace its path. The retracing nature of the burnt trajectory can also be seen as we draw a Rayleigh line connecting the upstream curve and corresponding downstream curve (represented by arrows). Also, the sign of the slope of the Rayleigh line changes after the maximum point.

We plot the corresponding SL GR curve in fig(4(a)). The range of the downstream curve increases for GR case as compared to the SR case. Also initially the slope of the RL is larger for the GR shocks. The maximum of the pressure also has a much higher value. We also show the corresponding TL GR CA curve in fig(4(b)). The CA range for TL GR shocks is much larger, however, the maximum of the pressure remains more or less the same. The relativistic and GR curve range may differ but all of them lies in the same downstream curve defined by the QM EoS. The maximum of the pressure becomes more clear if we plot pressure as a function of number density (fig(5)). For a given HM curve the corresponding QM pressure is lowest for relativistic shocks. For the GR shocks the corresponding pressure is much higher both for the SL and TL shocks. It is also clear that the maximum pressure for the relativistic case arrives at much lower density whereas for the GR case they appear at much larger densities. 

Till now we have discussed the conversion from pure HM to pure quark matter. However, There can be matter phase in QS where we have both hadrons and quarks which is also termed as mixed phase \citep{glen}. Therefore, we can have a hybrid star (HYS) which has HM at the outer region, mixed phase in the intermediate and in the core of the star and in some massive stars also a pure quark phase at the core. The corresponding pressure as a function of number density for HYS. The maximum pressure for the HYS is significantly smaller than the QS, indicating the pressure difference between the NS and HYS are not very large.

\begin{figure*}
\centering
\includegraphics[width = 3.30in,height=2.750in]{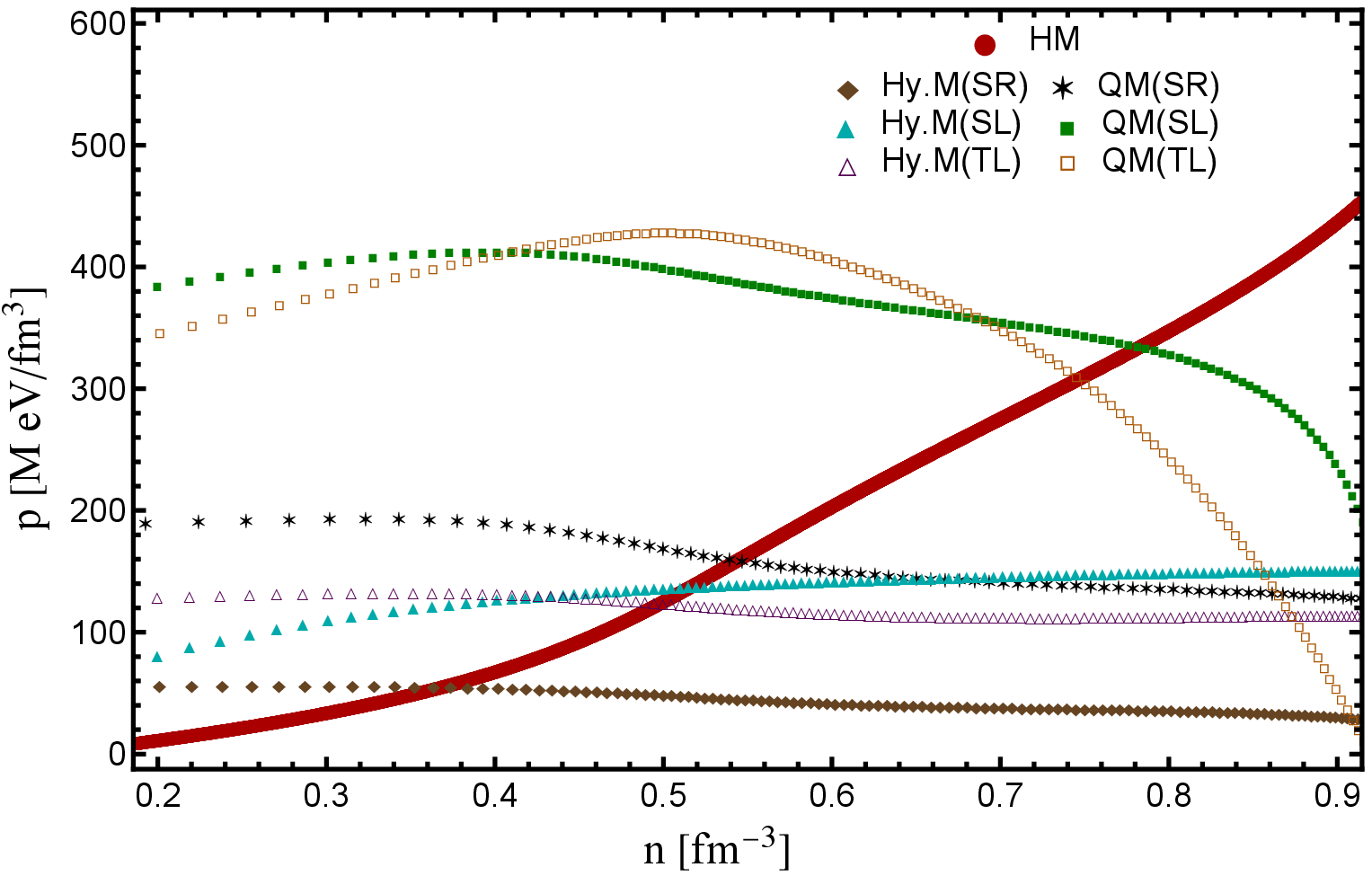}
\includegraphics[width = 3.30in,height=2.750in]{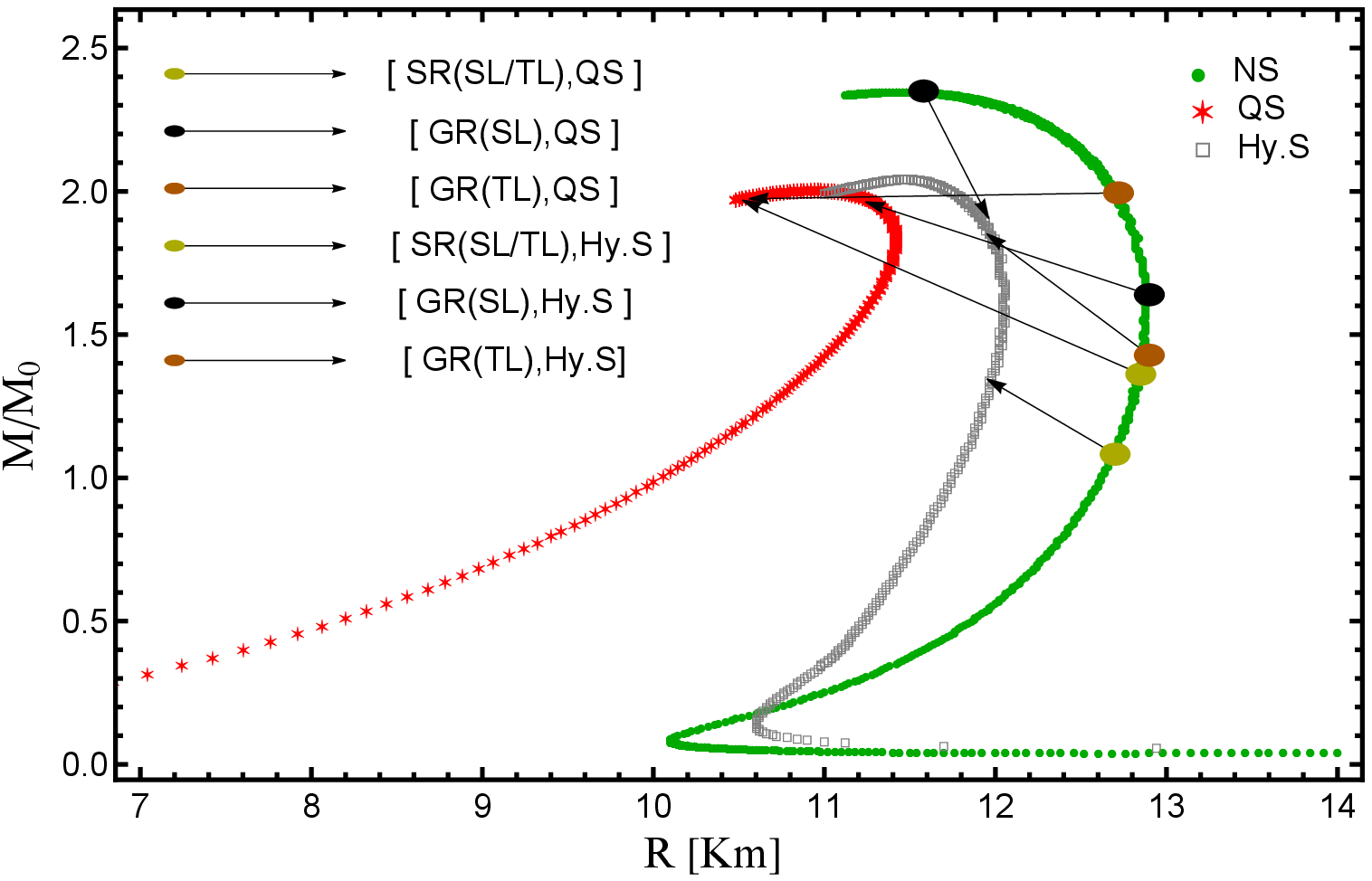}
\hspace{3.60cm} \scriptsize{(a)} \hspace{8.50cm} \scriptsize{(b)} 
\caption{(a) p as a function of baryon density (n) for HM and their corresponding downstream QM and Hy.M curves are shown in SR and GR (SL/TL) cases. The burnt matter pressure first rises and then decreases cutting their corresponding HM pressure curve at particular number density. (b) Mass-Radius relation for NS, QS and Hy.S are plotted. Each arrow indicate the maximum mass of phase transitioned QS (QM) and Hy.S (Hy.M) corresponding to input NS (HM) for both relativistic (SL/TL) and general relativistic (SL/TL) cases.}
\end{figure*}

It was earlier shown that the maximum pressure of the QM is used to calculate the maximum mass of the QS which results from the combustion of a parent NS \citep{ritam-irfan}. 
The occurrence of the maximum mass is shown in fig(6). The mass-radius diagram gives the sequences of stars masses with their corresponding radius. It also gives the maximum mass that can be attained for a given EoS. In the figure the green solid curve gives the mass-radius sequence of NS. The maximum mass it can reach is about 2.35 $M_\odot$  corresponding to a radius of about 12 km. The mass-radius sequence of QM EoS is shown with red curve which has a maximum mass of 2.05 $M_\odot$ corresponding to a radius of about 10.7 Km. However, if we assume that the quark star is obtained from the combustion of a parent NS (with relativistic calculation), then the maximum mass of the QS is about 1.963 $M_\odot$. The parent NS has a mass of about 1.56 $M_\odot$. Therefore, such combustion are highly unlikely as it needs external source of energy. However, the maximum masses of the daughter QS obtained from GR shocks are in agreement with maximum mass of the EoS sequence (as is shown in the figure). For GR SL shocks the mass difference is again huge (between the parent NS and daughter QS); however, for GR TL shocks the mass difference is negligible implyinga possible combustion process. 

However, if we now study the combustion from NS to HYS the situation is different. Again the maximum mass of duaghter QS is lower than the EoS sequence and is minimum for relativistic shocks. Relativistic shocks are unable to combust a NS and result a HYS more massive than $1.3 M_\odot$. The situation is better for GR shocks which can produce a HYS of mass as large as $1.9 M_\odot$. However, for SL shocks huge amount of external energy is required to initiate such combustion but for TL shocks NS can undergo exothermic combustion and produce HYS. Remember, there can be another scenario where a NS undergoing combustion to either QS or HYS can become unstable and collpse to a Black hole.



\subsection{Matter velocities across the front}

The matter velocities across the front are a valuable tool to understand the properties of shock induced combustion. Combustion can be either detonation (fast burning where the combustion and the shock front almost coincides) or a deflagration (slow burning). If the velocity of the burned matter is larger than unburnt matter then phase transition corresponds to the detonation, whereas if its speed is smaller than unburnt matter then phase transition resembles the deflagration or slow combustion. This is expressed as
\begin{align}
v_{upstream}>v_{downstream} \Rightarrow Deflagration \nonumber \\
v_{upstream}<v_{downstream} \Rightarrow Detonation   \nonumber
\end{align}

\begin{figure*}
\centering
\includegraphics[width = 3.30in,height=2.750in]{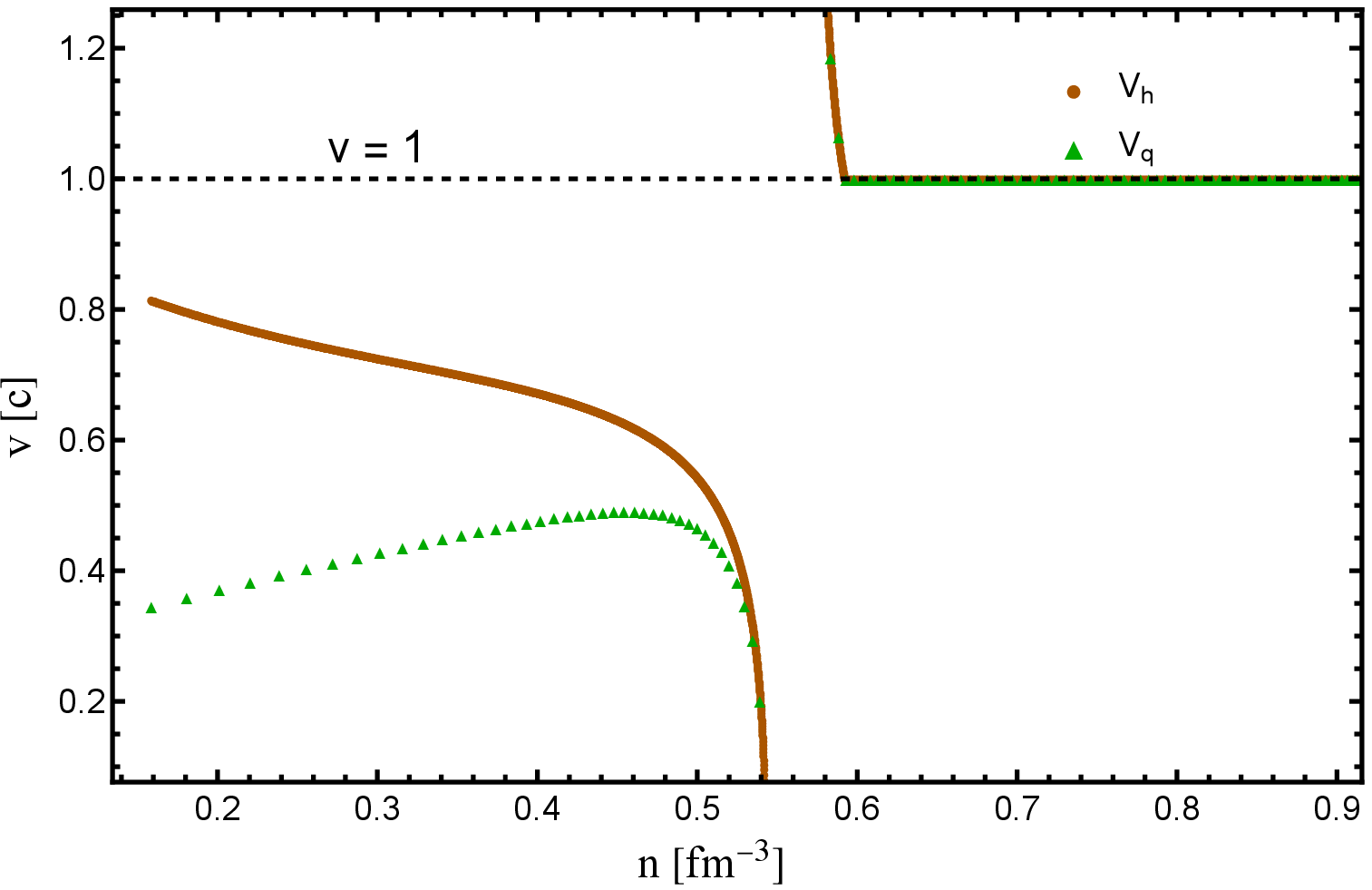}
\includegraphics[width = 3.30in,height=2.750in]{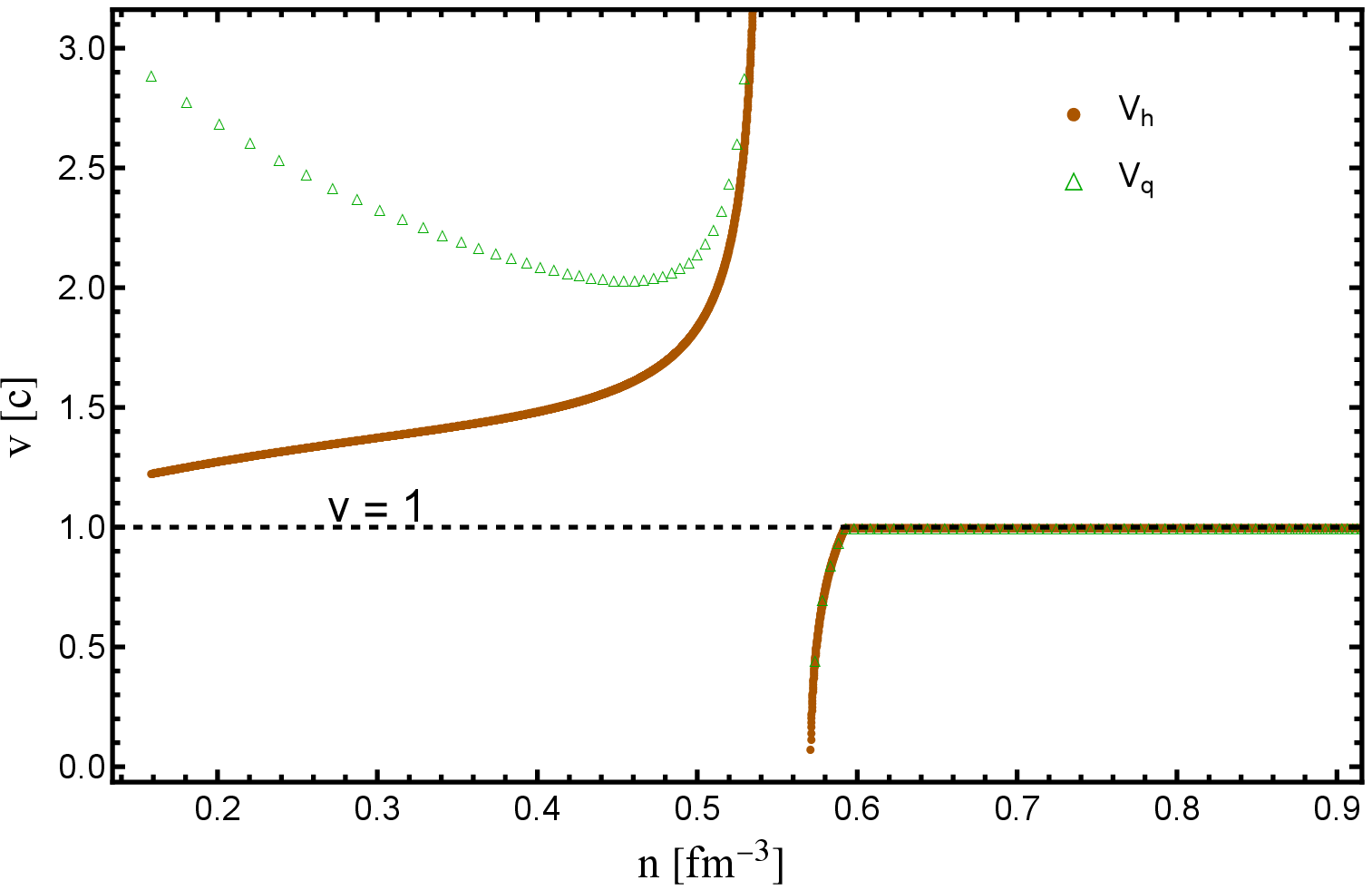}
\hspace{3.60cm} \scriptsize{(a)} \hspace{8.50cm} \scriptsize{(b)} 
\caption{The upstream ($v_a = v_h$) for HM EoS (red-circle)  and downstream ($v_b = v_q$) for QM EoS (green-triangle) velocities are shown as a function of number density for (a) SL and (b) TL relativistic shocks. For SL shocks, $v_h$ is always greater than $v_q$ at low density and both are subluminal but they tends to superluminal as density become much higher. However, for TL shocks, $v_q$ is less than $v_h$ at low density and are superluminal while they tends to superluminal as density become much higher.}
\end{figure*}

\begin{figure*}
\centering
\includegraphics[width = 3.30in,height=2.750in]{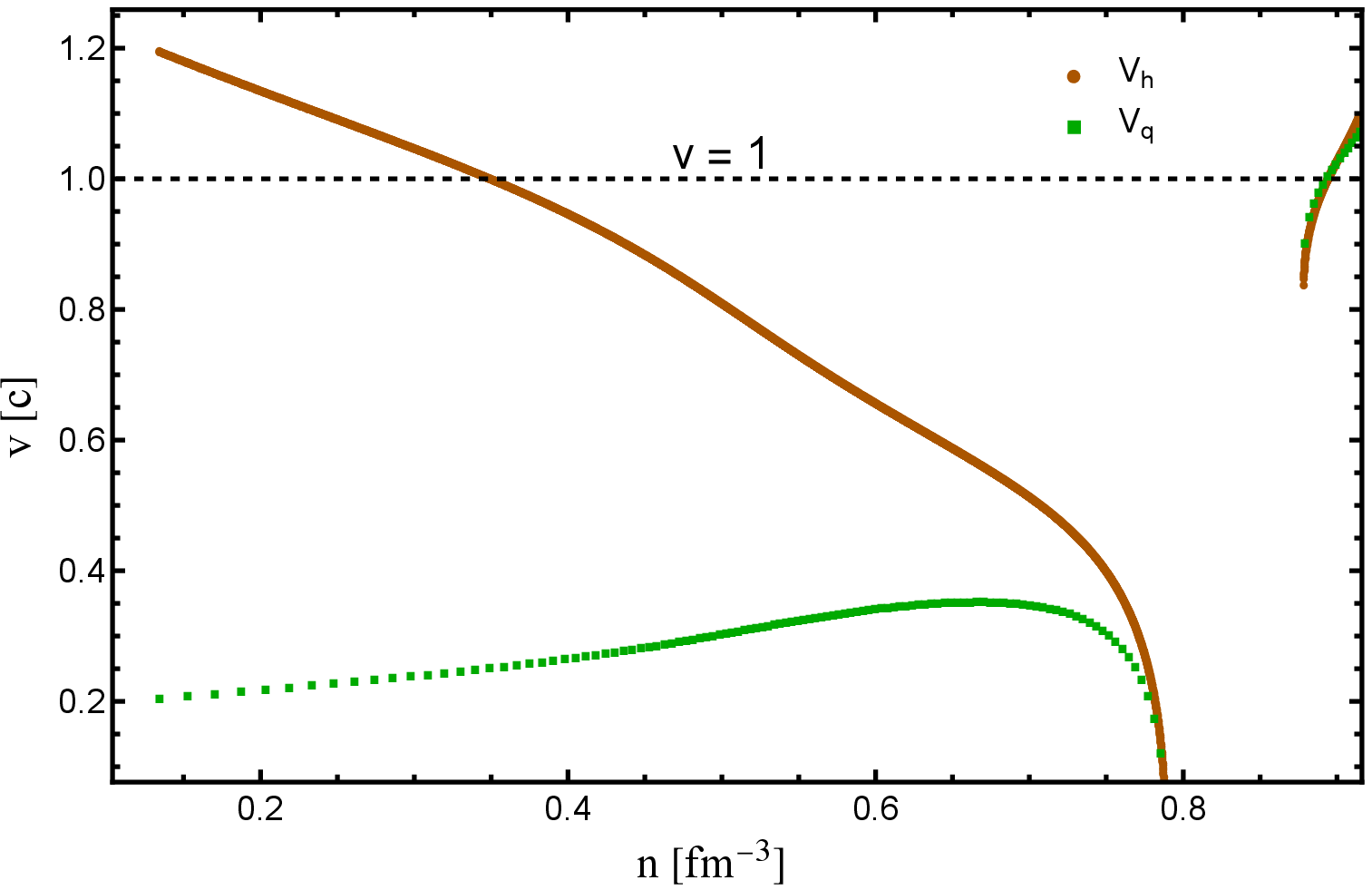}
\includegraphics[width = 3.30in,height=2.750in]{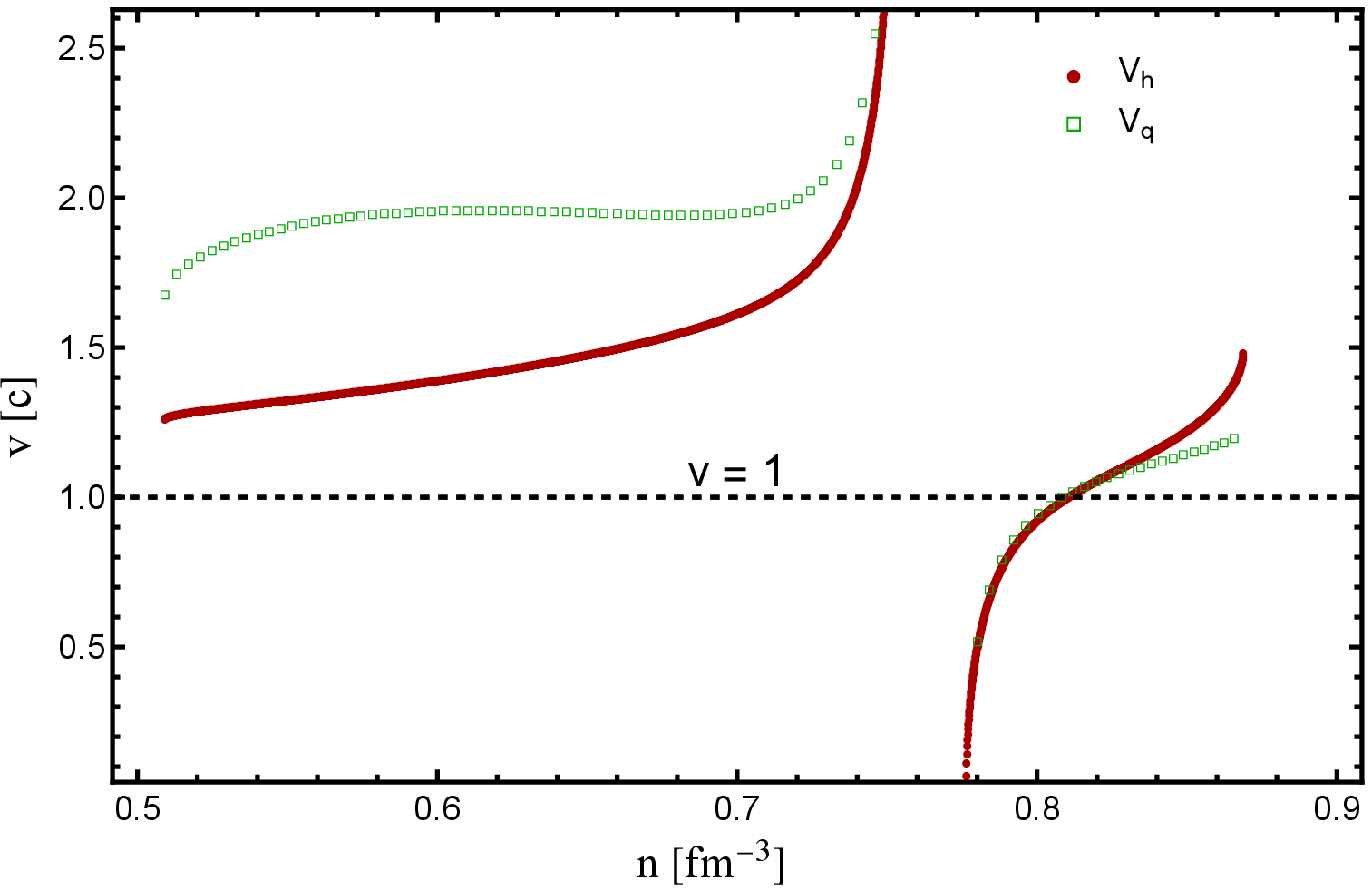}
\hspace{3.60cm} \scriptsize{(a)} \hspace{8.50cm} \scriptsize{(b)} 
\caption{The upstream ($v_a = v_h$) for HM EoS (red-circle)  and downstream ($v_b = v_q$) for QM EoS (green-square) velocities are shown as a function of number density for (a) SL and (b) TL general relativistic shocks. For SL shocks, $v_h$ become superluminal at much less densities and is always greater than $v_q$ (subluminal). However, for TL shocks, $v_q$ is less than $v_h$ at low density but each are superluminal, while both tends to superluminal as density become much higher.}
\end{figure*}

\begin{figure*}
\centering
\includegraphics[width = 3.30in,height=2.750in]{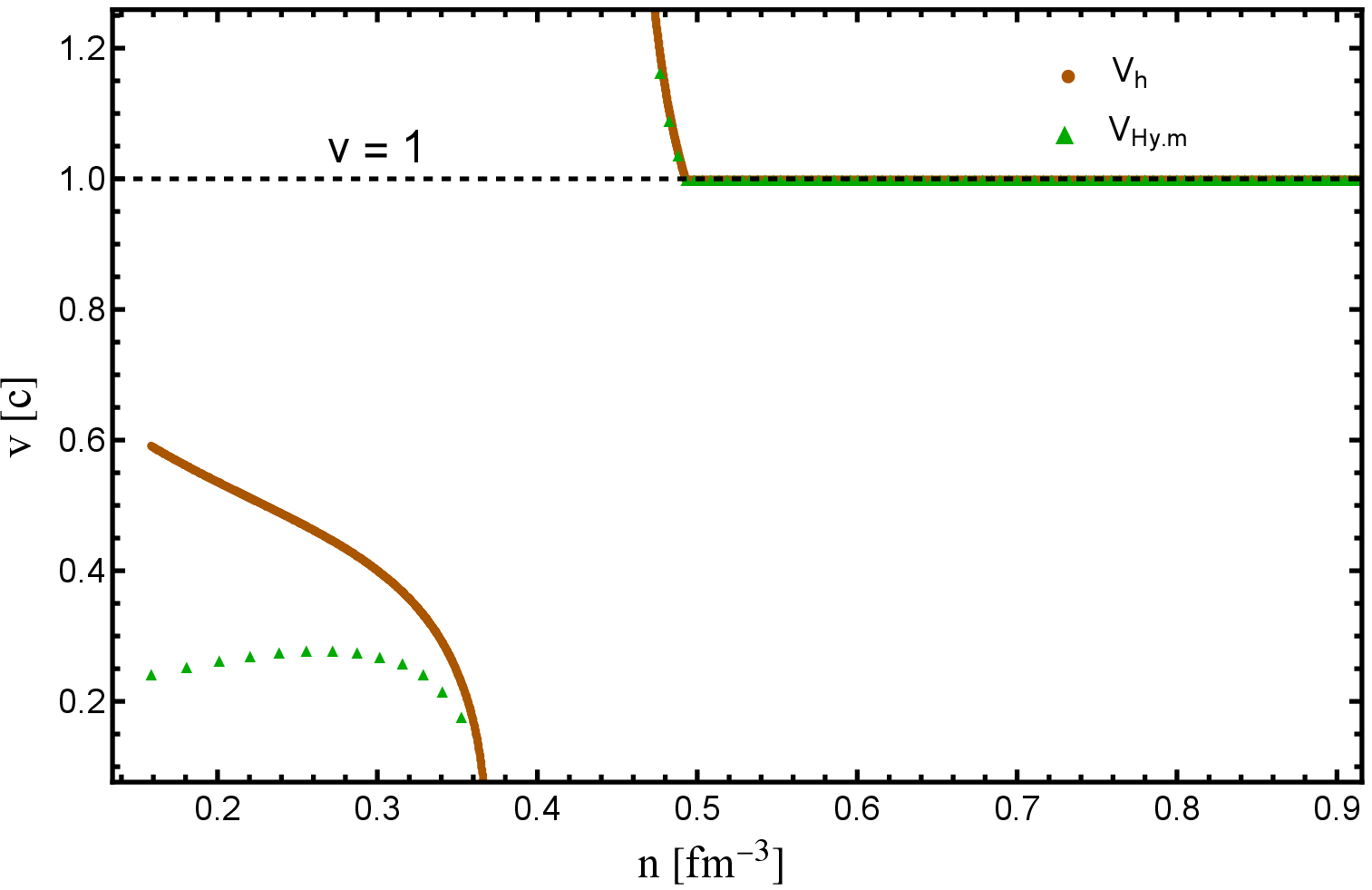}
\includegraphics[width = 3.30in,height=2.750in]{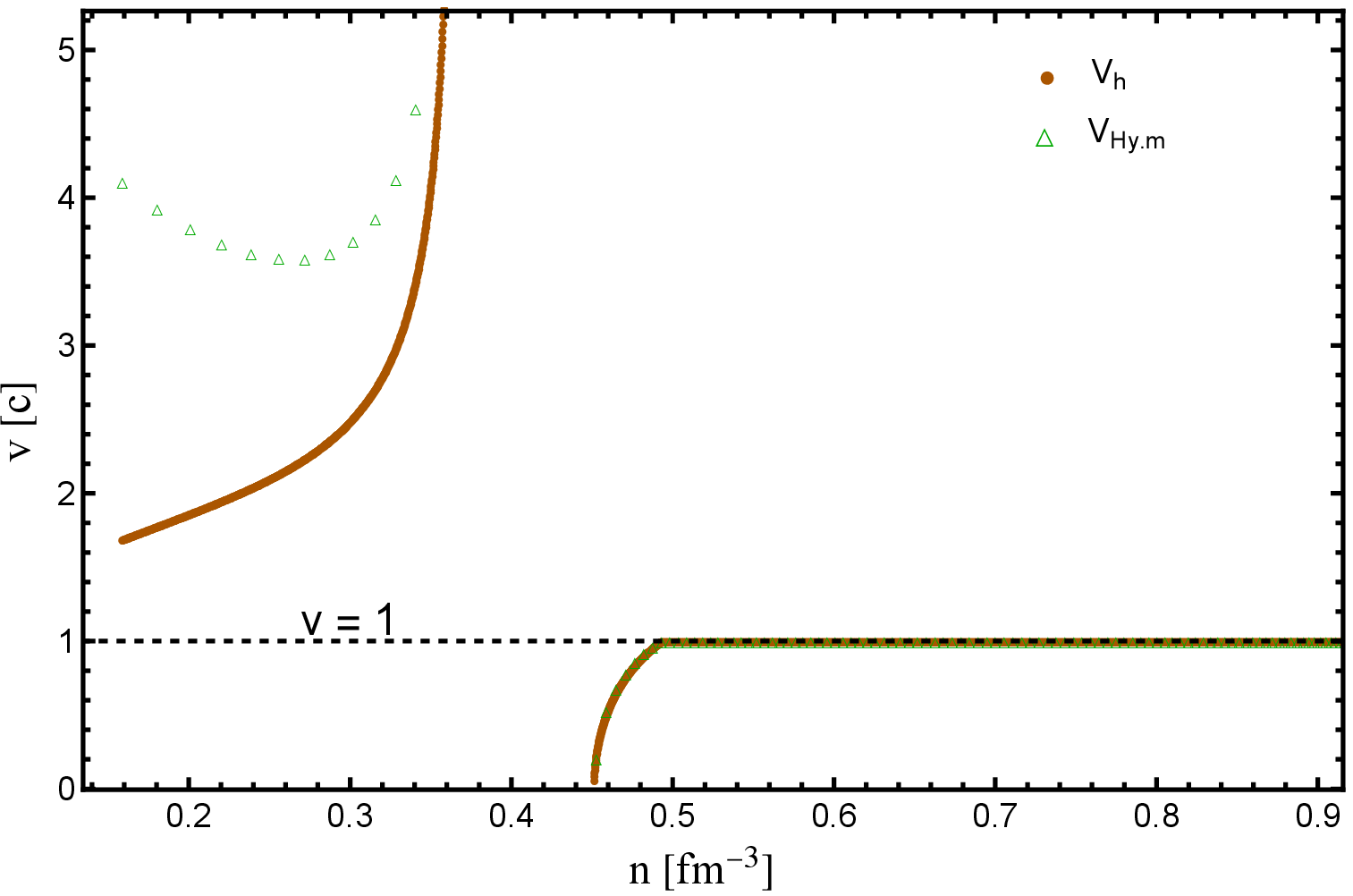}
\hspace{3.60cm} \scriptsize{(a)} \hspace{8.50cm} \scriptsize{(b)} 
\caption{The upstream ($v_a = v_h$) for HM EoS (red-circle)  and downstream ($v_b = v_{Hy.M}$) for Hy.M EoS (green-triangle) velocities are shown as a function of number density for (a) SL and (b) TL relativistic shocks.  For SL shocks, $v_h$ is always greater than $v_{Hy.M}$ at low density and both are subluminal but they tends to superluminal as density become much higher. However, for TL shocks, $v_{Hy.M}$ is less than $v_h$ at low density and are superluminal while they tends to superluminal as density become much higher.}
\end{figure*}

\begin{figure*}
\centering
\includegraphics[width = 3.30in,height=2.750in]{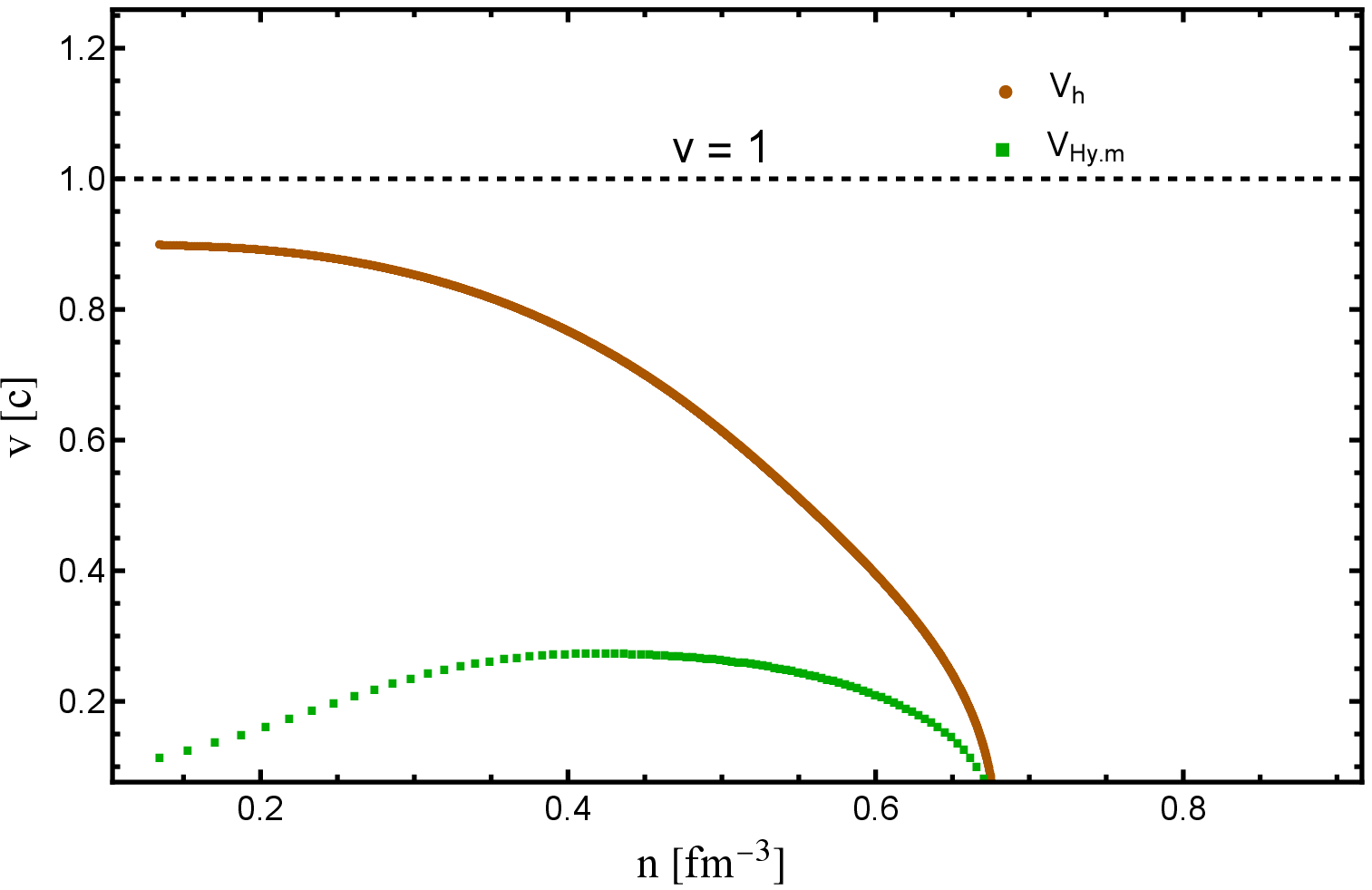}
\includegraphics[width = 3.30in,height=2.750in]{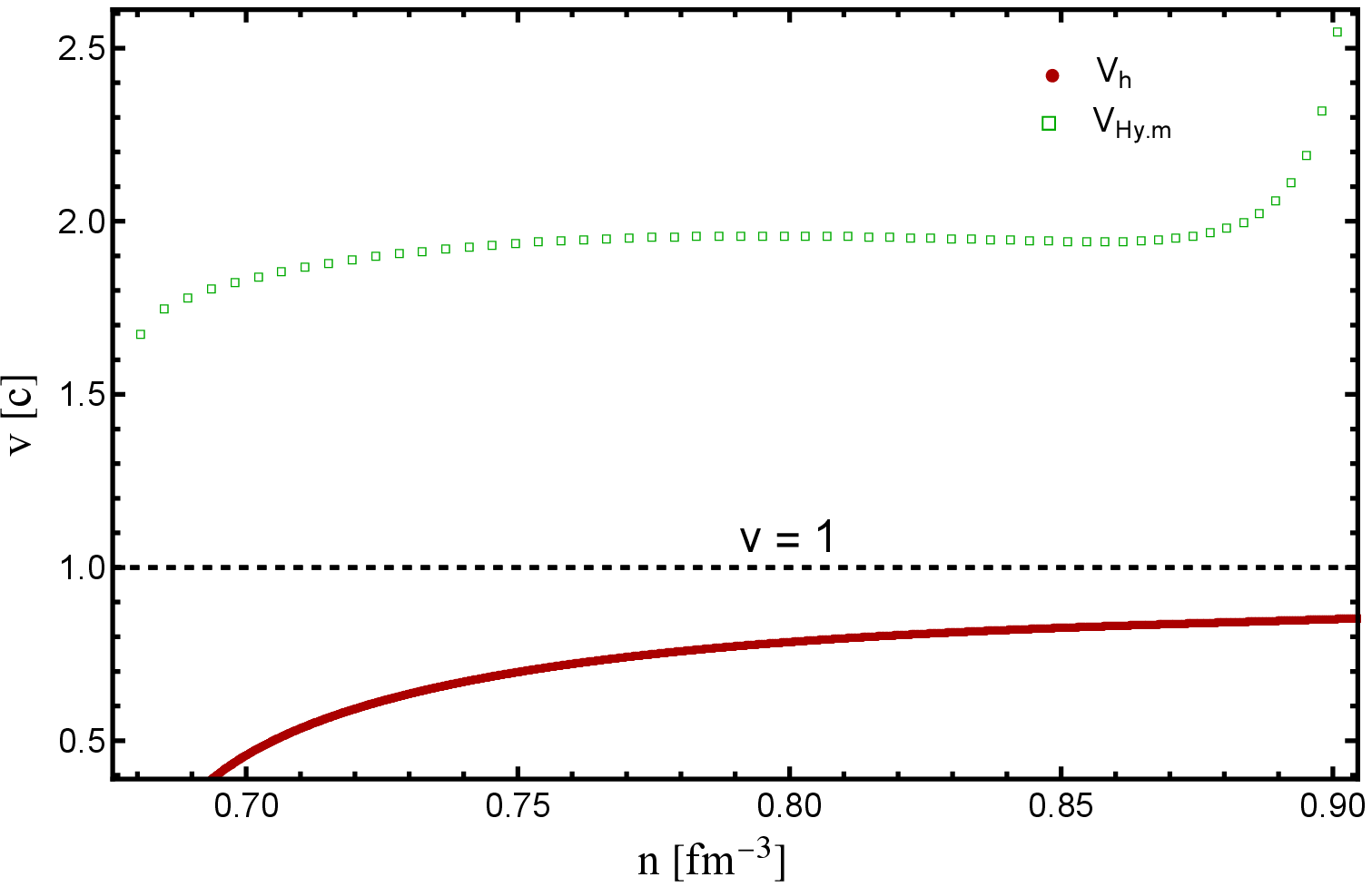}
\hspace{3.60cm} \scriptsize{(a)} \hspace{8.50cm} \scriptsize{(b)} 
\caption{The upstream ($v_a = v_h$) for HM EoS (red-circle)  and downstream ($v_b = v_{Hy.M}$) for Hy.M EoS (green-square) velocities are shown as a function of number density for (a) SL and (b) TL general relativistic shocks. For SL shocks, $v_h$ is always greater than $v_{Hy.M}$ and both are always subluminal. However, for TL shocks, $v_{Hy.M}$ is much larger than $v_h$ in which $v_q$ is superluminal and $v_h$ is subluminal always.}
\end{figure*}

We have shown the matter velocities for relativistic SL and TL shocks in fig(6) for combustion from HM to QM. The matter velocity of the hadronic matter is shown by the red curve whereas the quark matter velocity is shown in green. For the SL shocks we find that at low densities hadronic matter velocity is much higher than quark matter velocity indicating a deflagration combustion. As the density increases their difference decreases and at a density $0.54$ fm$^{-3}$ the becomes equal and goes to zero indicating no combustion process. This indicates that at such densities the shock induced combustion from HM to QM is not possible. At about $0.58$ fm$^{-3}$ matter velocities again becomes non zero and attains a velocity close to that of velocity of light. The properties of matter velocities for relativistic TL shocks are just the inverse of the SL shocks. Therefore, at low densities quark matter velocities are greater than the hadronic matter velocities but both are greater than the speed of light, indicating almost instantaneous combustion. At higher densities they attain a velocity close to that of speed of light.

However, the situation is quite different for GR shocks as shown in fig(7). Although, at low density $v_q$ is less than $v_h$ (just like the relativistic shocks), the value of $v_h$ at low densities is greater than the speed of light. This situation continues till very large density values of about $0.8$ fm$^{-3}$. Beyond that matter velocities becomes zero indicating that the combustion process is not sustainable at such densities. At much higher densities (which is usually not realized even at NS cores) matter velocities can again be finite and even greater than the speed of light. For the TL GR shocks initially $v_q$ is greater than $v_h$ and both are greater than the speed of light indicating instantaneous detonation. This continues till $0.75$  fm$^{-3}$ where the velocities becomes infinitely large. Beyond that the velocities becomes zero. However, at about $0.78$ fm$^{-3}$ they becomes finite a less than the speed of light initially, but later very high densities again they become greater than the speed of light.

Similar velocity plot for relativistic and GR shocks combustion from NS to HYS are shown in fig (8 \& 9). The subluminal velocity for relativistic SL shocks appears only at lower densities having a deflagration type of combustion. At higher densities matter velocities are either luminal or super-luminal. As expected the matter velocities are just the opposite for relativistic TL shocks. The scenario is quite different for GR shocks. The velocities are subluminal for quite an extended range having an deflagration combustion for GR SL shocks. Beyond that the velocities are imaginary indicating no combustion for massive stars. For the TL shocks the matter velocities are imaginary at low densities whereas at high densities the quark matter velocity is super-luminal whereas the hadronic matter velocity is sub-luminal indicating detonation type of combustion only for very massive stars. It should be mentioned that although matter velocities are super-luminal for quite a range of densities the front velocity is almost sub-luminal for most of the density range.

\section{Summary and Discussion}

Study of Shock wave are important to understand many physical scenarios and particle acceleartion in astrophysics. One needs relativistic calculation to understand physical phenomena at such regimes.
The relativistic shock condition was done by Taub a long ago \citep{taub}, however, he considered only the SL shocks. Csernai \citep{csernai} introduced the idea of TL shocks which can also in principle exist. Although, few GR shock calculation exists in literature a detailed analysis of SL and TL shock was lacking. In the present work we present a detail analysis of GR SL and TL shocks and compare them with the relativistic case to observe the difference. We study the GR shock in the NS system where we assume that the shock induced transition deconfines HM to QM. 

We initially derive the RH condition for both the SL and TL GR shocks and from there we derive the CA equation. However, unlike the SR shocks where the CA equation was same for SL and TL case, the CA equation for GR SL and TL shocks are different because of the occurrence of the metric potential. Also, the matter velocities for relativistic SL and TL shocks were inverse of each other which is not the case for GR shocks. We also find that if the gravitational potentials becomes either zero or same on either side of the shock front the GR conditions reduces to SR conditions.

Assuming a initial state for the HM we solve the CA equation to get the final state of the QM in the star. The matter velocities can then be calculated from the thermodynamical variables of the initial and final states. With a given initial state we can study the matter velocities to check for which final state the matter velocities becomes imaginary or breaks the speed of light limit. It was interesting to find that even with ensuring the causality condition there are certain final state (with a given initial state) where the matter velocities can be super-luminous. 

For the relativistic shocks, given an initial state the region with velocities greater than the speed of light are just the opposite for SL and TL shock in the $p-\epsilon$ plane. However, the picture for GR shocks are quiet different. For a given initial state, the region of imaginary velocities and region of velocities greater than speed of light are also function of the mass of the star (which determines the gravitational potentials). Also, the regions for SL and TL shocks are not complimentary. The occurrence of regions where matter velocities becomes greater than the speed of light is of great importance because if we are able to reach such final states we can have almost super-luminous combustion/shocks which can important astrophysical implications.

The GR shock calculation is extended to study the combustion of HM to QM in NS and to calculate the maximum mass of the daughter QS which results from the combustion of parent NS. Previous SR calculation indicated that the maximum mass of the PT quark star is much less than that of the actual maximum mass of the quark sequence; however, the GR calculation shows that the combusted QS maximum mass is in agreement with the maximum mass of the QS EoS sequence. The GR calculation shows that combustion from NS to QS is more viable with TL shocks. The combustion of massive NS to HYS with relativistic calculation is almost impossible however, GR calculation can have massive NS combusting to massive HYS.

The velocity of the matter phases on either side of the front can be examined to determine the combustion process where we find that both for the SR and GR SL shock combustion at low density involves deflagration process. For the TL shocks at low density the velocities are higher than the speed of light and can signify super-luminous combustion.

It should be mentioned that the calculation is done from the rest frame of the front. Although, the matter velocities can be greater than the speed of light the front velocity for most of the case is sub-luminous. We are in the process of obtaining the dynamical equation s for SL and TL GR shocks which would give us a clear picture about the shock propagation. However, this analysis shows that the TL shock velocities are in principle greater than SL shock velocities. Most of the shock calculation in astrophysical scenario are studied with SL shocks and therefore, it will be interesting to check them for TL shocks. The GR shocks are dependent of the mass of the system and this is another aspect that can be studied in more detail where large masses (like black hole) can be considered.

\section{Acknowledgments}
The authors are grateful to Indian Institute of Science Education and Research Bhopal for providing all the research and infrastructure facilities.

\end{document}